%% file: main.tex
\documentclass[10pt,twocolumn,letterpaper]{article}
\usepackage{iccv}
\usepackage[pdftex]{graphicx} 
\usepackage{times}
\usepackage{epsfig}
\usepackage{amsmath}
\usepackage{amssymb}
\usepackage{hyphenat}
\usepackage{multirow}
\usepackage[numbers]{natbib}
\usepackage{paralist}
\usepackage{subcaption}
\usepackage{makecell}
\usepackage{mwe}

\usepackage[pagebackref=true,breaklinks=true,letterpaper=true,colorlinks,bookmarks=false]{hyperref}

\iccvfinalcopy 
\ificcvfinal\fi

\include{src/headers.tex}

\begin{document}
\title{Weakly Supervised Caveline Detection For AUV Navigation Inside \\ Underwater Caves}  % **** Enter the paper title here

\author{Boxiao Yu$^{1}$, Reagan Tibbetts$^{2}$, Titon Barua$^{2}$, Ailani Morales$^{1}$, Ioannis Rekleitis$^{2}$ and Md Jahidul Islam$^{1}$ \\
{\tt\small \{boxiao.yu, ailanimorales\}@ufl.edu, jahid@ece.ufl.edu, \{rbt,baruat\}@email.sc.edu,  yiannisr@cse.sc.edu} \\
{
\small $^{1}$RoboPI Laboratory, Department of ECE: University of Florida, FL 32611, USA.} \\ 
{
\small $^{2}$AFRL Laboratory, Department of CSE: University of South Carolina, SC 29208, USA. }
\thanks{This research has been supported in part by the National Science Foundation under grants 1943205, 1919647, 2024741, 2024541 and 2024653. The authors would also like to acknowledge the help of the Woodville Karst Plain Project (WKPP), El Centro Investigador del Sistema Acuífero de Quintana Roo A.C. (CINDAQ), and Ricardo Constantino, Project Baseline in collecting data, providing access to challenging underwater caves, and mentoring us in underwater cave exploration.}
\thanks{This pre-print is accepted for publication at the IROS 2023. [06/23]}
}

\maketitle
\thispagestyle{empty}

%%%%%%%%% Abstract
\begin{abstract}
\textit{Underwater caves are challenging environments that are crucial for water resource management, and for our understanding of hydro\hyp geology and history. Mapping underwater caves is a time-consuming, labor-intensive, and hazardous operation. For autonomous cave mapping by underwater robots, the major challenge lies in vision-based estimation in the complete absence of ambient light, which results in constantly moving shadows due to the motion of the camera\hyp light setup. Thus, detecting and following the \emph{caveline} as navigation guidance is paramount for robots in autonomous cave mapping missions. In this paper, we present a computationally light caveline detection model based on a novel Vision Transformer (ViT)-based learning pipeline. We address the problem of scarce annotated training data by a weakly supervised formulation where the learning is reinforced through a series of noisy predictions from intermediate sub-optimal models. We validate the utility and effectiveness of such weak supervision for caveline detection and tracking in three different cave locations: USA, Mexico, and Spain. Experimental results demonstrate that our proposed model, \emph{CL-ViT}, balances the robustness-efficiency trade-off, ensuring good generalization performance while offering 10+ FPS on single-board (Jetson TX2) devices}.
\end{abstract}

%%%%%%%%% BODY TEXT - ENTER YOUR RESPONSE BELOW
\input{src/Intro.tex}

\input{src/Related_Work.tex}

\input{src/Methodology.tex}

\input{src/Experiments.tex}
\input{src/Conclusion.tex}

%-------------------------------------------------------------------------

{\small
\bibliographystyle{ieee}
\bibliography{main}
}

\end{document}

%% file: src/headers.tex
\usepackage{subcaption}
\usepackage{amsmath} % assumes amsmath package installed
\usepackage{amssymb}  % assumes amsmath package installed
\usepackage{csquotes}

\usepackage{array}
\usepackage{enumerate}
\usepackage{multirow} 
\usepackage{float}
\usepackage{paralist}
\usepackage{amsfonts}
\usepackage{longtable}

\usepackage{xspace}

\newcolumntype{L}[1]{>{\raggedright\let\newline\\\arraybackslash\hspace{0pt}}m{#1}}
\newcolumntype{C}[1]{>{\centering\let\newline\\\arraybackslash\hspace{0pt}}m{#1}}
\newcolumntype{R}[1]{>{\raggedleft\let\newline\\\arraybackslash\hspace{0pt}}m{#1}}
\graphicspath{ {img/} }

\newcommand{\nostarnote}[1]{}
\newcommand{\baad}[1]{} 
\usepackage[size=small]{caption}
\usepackage{color}

\usepackage{makecell}

\usepackage{color}

%% file: src/Intro.tex
\section{Introduction}
\begin{figure}[t]
     \centering
     {\includegraphics[width=0.98\columnwidth]{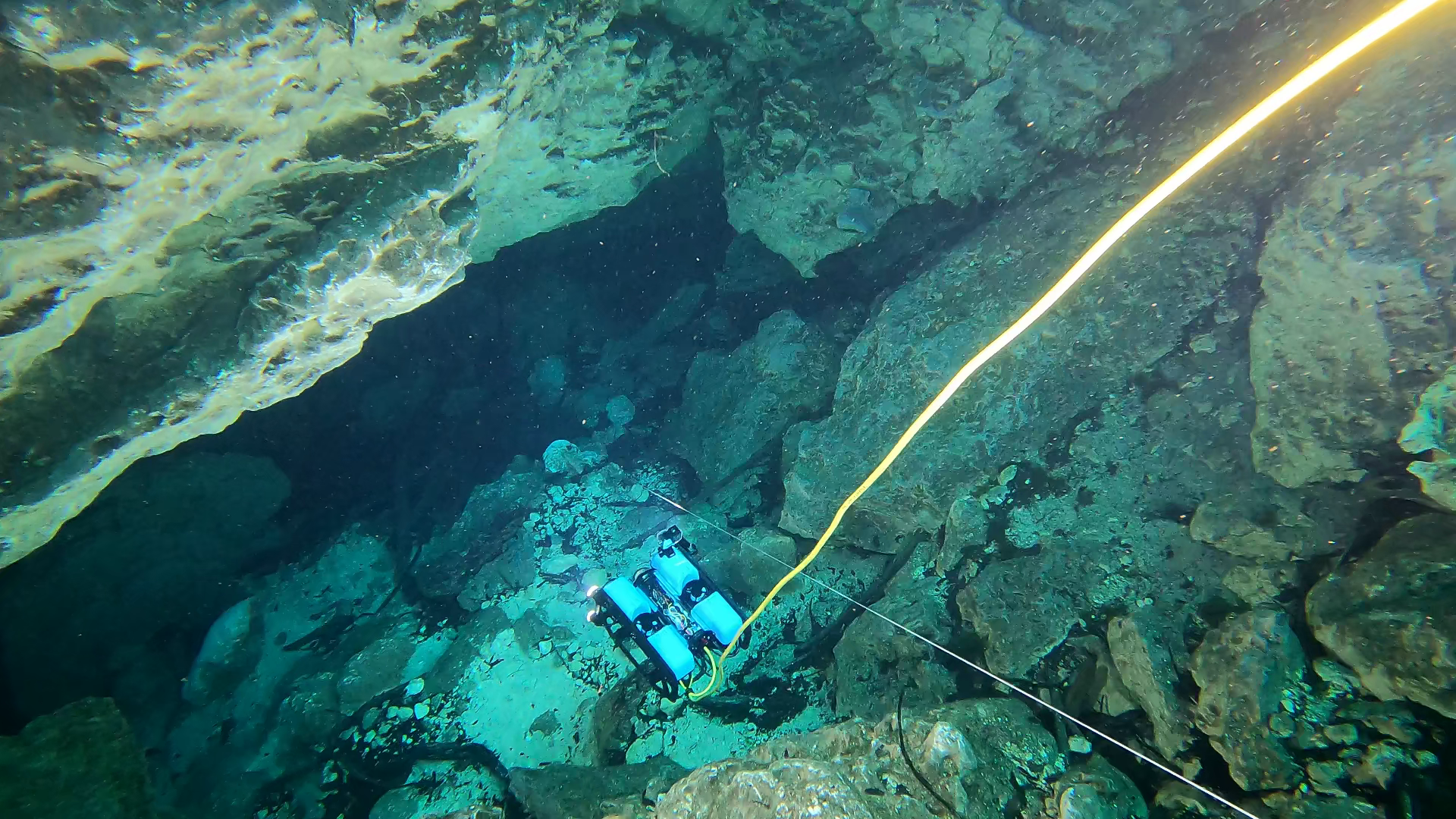}}
     {\includegraphics[width=0.98\columnwidth]{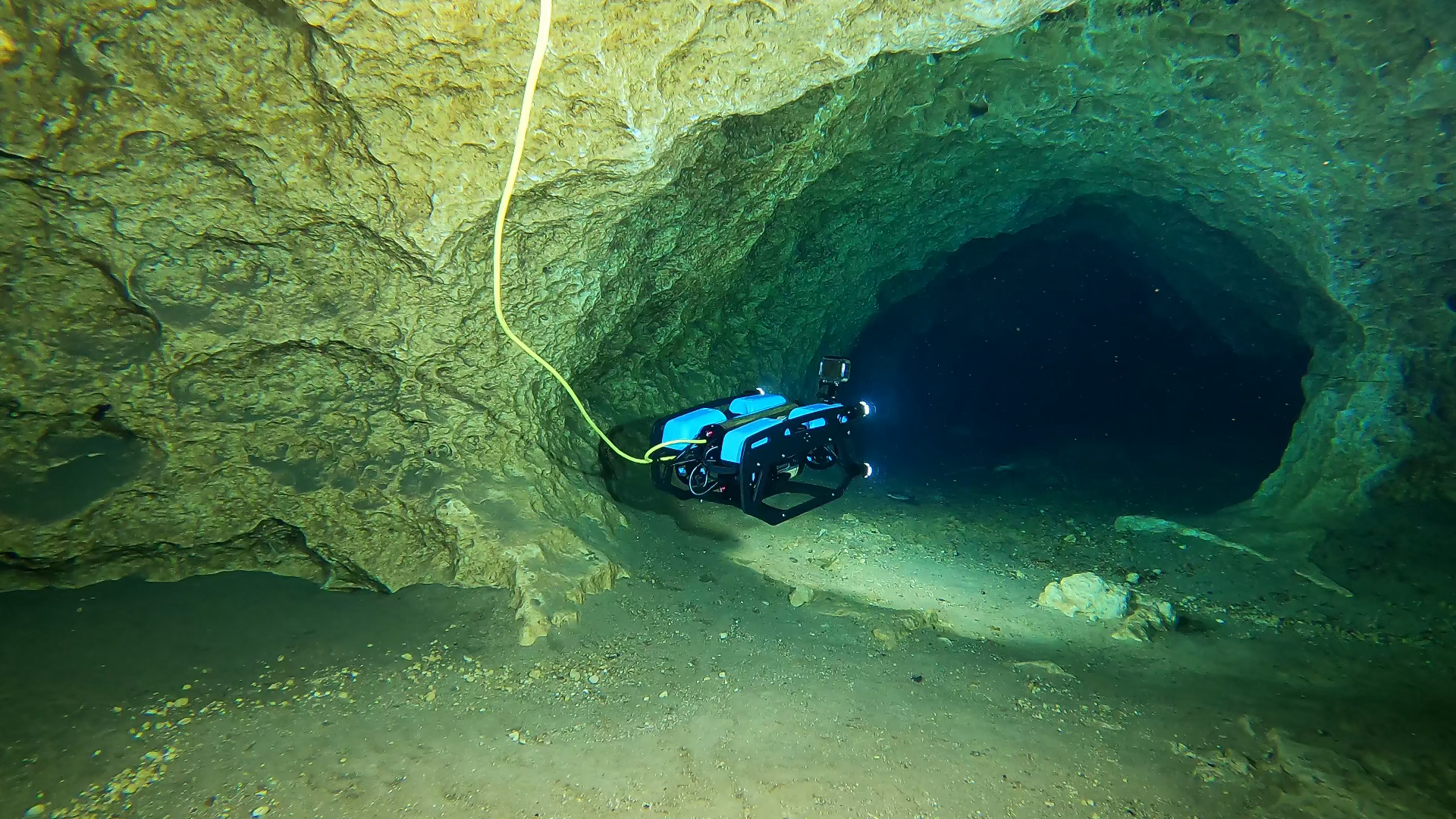}}
     \caption{A BlueROV2 operating inside the cave, Orange Grove Sink, Florida, USA. Note that the umbilical is connecting the ROV to a surface operator.}%
     \vspace{-5mm}
     \label{fig:beauty_shot}
 \end{figure}

Underwater caves play a crucial role in monitoring and tracking groundwater flows in Karst topographies, while almost $25\%$ of the world's population relies on Karst freshwater resources~\cite{karstbook}. Moreover, underwater caves often present a pristine \emph{capsule} preserved in time with major archaeological secrets~\cite{gonzalez2008arrival}. Underwater cave exploration and mapping by human divers, however, is a tedious, labor-intensive, extremely dangerous operation even for highly skilled people~\cite{buzzacott2009american}. Therefore, enabling Autonomous Underwater Vehicles (AUVs) and Remotely Operated Vehicles (ROVs) to enter, navigate, map, and finally exit an underwater cave is important to ensure the safety and efficacy of a mapping mission, as well as to potentially generate more accurate maps; Fig.~\ref{fig:beauty_shot} shows an ROV deployment scenario inside the Ballroom cavern at Ginnie Springs, Florida. 

The first cardinal rule of cave diving as set by Sheck Exley is \emph{``Always use a single, continuous guideline from the entrance of the cave throughout the dive.''}~\cite{exley1986basic}. Such guidelines, from here on termed {\em cavelines}, exist in all explored underwater caves and they provide the skeleton of the main passages. Mapping underwater caves is a multi\hyp layered process. When a new section of a cave is discovered, a caveline is set identifying the passage. Consequently, the caveline is surveyed marking the depth and orientation at the points where the line is attached to the cave (floor, ceiling, or walls) and the distance between attachment points (called placements). These surveys produce a one-dimensional retraction of the three-dimensional environment. Recording all this information together with additional observations~\cite{Burge1988Survey} such as distance to the walls, ceiling, and floor-- is a challenging, time\hyp consuming, and error\hyp prone process.

\begin{figure}[h]
     \centering
     {\includegraphics[width=\columnwidth]{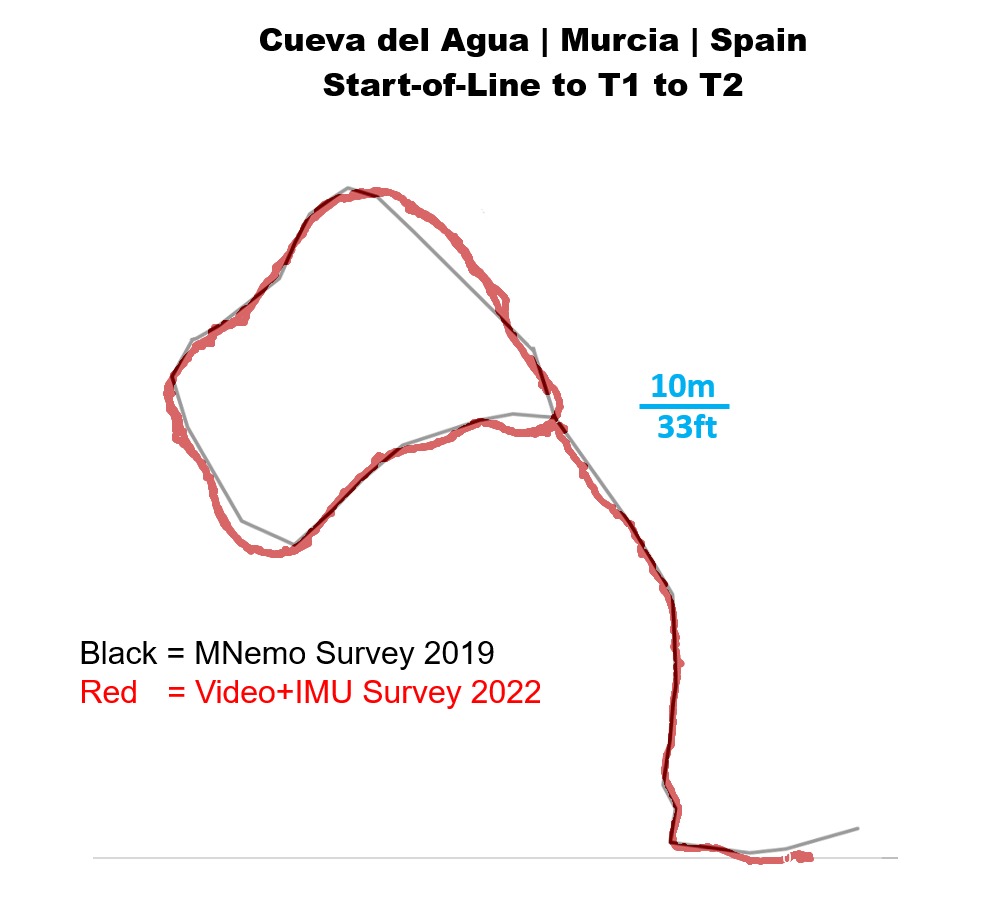}}%
     \vspace{-5mm}
     \caption{Estimated trajectory together with manually measured ground truth from baseline; Cueva Del Agua, Spain.}
     \label{fig:Spain}
 \end{figure}

Our earlier work~\cite{JoshiICRA2022} utilizing a GoPro-9 action camera resulted in high-precision camera trajectory estimation by a Visual-Inertial Odometry (VIO) algorithm~\cite{RahmanIJRR2022}, which is comparable to manually surveyed caveline (see Fig.~\ref{fig:Spain}). Moreover, the collected data are continuous spatiotemporal videos, which we used to generate weakly labelled data, and demonstrated that iterative filtering of mislabelled samples can help regulate sample extraction for improved learning~\cite{ModasshirICRA2020}. We found that the major challenges of data-driven solutions for problems such as the caveline detection and tracking, are: ($i$) learning from very few annotated samples; and ($ii$) ensuring generalization performance across different waterbodies, scene geometries, and optical degradations.

In this work, we address the aforementioned issues by developing a weakly supervised Vision Transformer (ViT)-based learning pipeline for autonomous caveline detection by AUVs. We demonstrate that with a limited amount of annotated training samples, learning can be reinforced iteratively from intermediate sub-optimal solutions. Specifically, after each \textit{training phase}, the \emph{weak} predictions are carefully sorted by a human expert into positive (\ie, accurate) and negative (erroneous) labels. The noisy positive samples and a fraction of newly annotated negative samples are then fused to reinforce learning in the subsequent phases. With a series of experiments, we show that robust caveline detection with good generalization performance can be achieved with only $1.5$K-$2$K annotated samples within $2$-$3$ training phases. 

We conduct experiments on \textbf{three cave systems} in different geographical locations: the Devil's system in Florida, USA; Dos Ojos Cenote, QR, Mexico; and Cueva del Agua in Murcia, Spain. We compile the data into three sets containing different types of cavelines, in terms of thickness and color, and different background and optical degradation levels. In order to ensure robustness, we evaluate both intra-set and inter-set detection performance -- with the goal of achieving good generalization performance on data from an unseen location. We validated the effectiveness of our weakly supervised multi-phase training for several genres of prominent state-of-the-art (SOTA) models based on convolutional neural networks (CNNs), attention networks, conditional random fields (CRFs), U-shaped encoder-decoders, and ViTs.    

Moreover, we develop a novel ViT-based learning pipeline named \textbf{CL-ViT} that offers the design choice of a \emph{base model} with EfficientNetB5~\cite{tan2019efficientnet} backbone and a \emph{light model} with MobileNetV3~\cite{howard2019searching} backbone for offline use (by surface operators) and online processing (onboard AUVs), respectively. The base model surpasses SOTA detection performance and provides fine-grained caveline localization in image space. Additionally, with a highly efficient MobilenetV3 backbone~\cite{howard2019searching} CL-ViT light model has only $12.67$M parameters, which is about $51.55\%$ less than DeepLabv3+~\cite{chen2018encoder} and $46.61\%$ less than PAN~\cite{li2018pyramid} -- two of the best competitor baselines. As a result, it offers significantly faster inference rates: over $215.79$ FPS on an NVIDIA\texttrademark~RTX $3060$ and $10.71$ FPS on a single-board Jetson TX2. 
A series of challenging test experiments reveal that CL-ViT offers consistent performance for detecting cavelines with the presence of shadow, lighting variations, and other optical artifacts. 
%Specifically, the performance margins of CL-ViT are within $n\%$ of the SOTA scores, while it has $*.*\%$-$51.55\%$ fewer parameters offering $7.5$-$8.4$ times faster inference.
\begin{figure*}[t]
\footnotesize
    \centering
    \begin{tabular}{ccc}
        \includegraphics[width=0.31\textwidth]{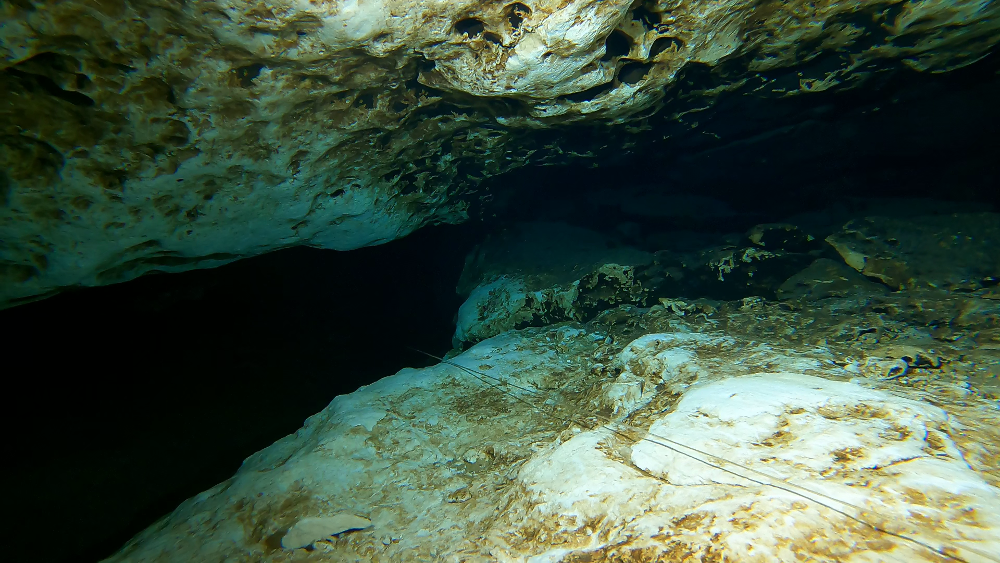}&
        \includegraphics[width=0.31\textwidth]{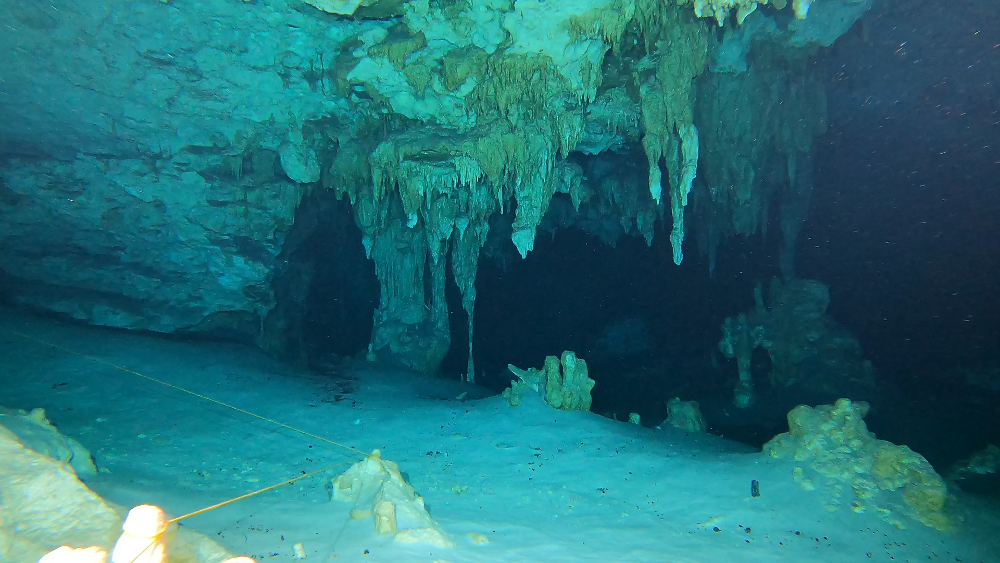}&
        \includegraphics[width=0.31\textwidth]{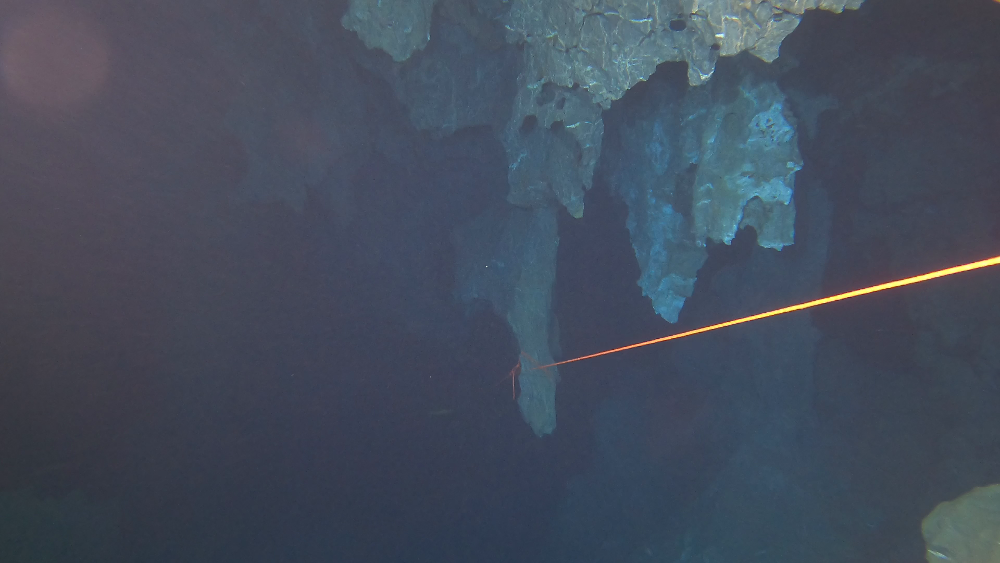}\\
        (a) Devil system, FL, USA & (b) Dos Ojos Cenote, QR, Mexico &(c) Cueva del Agua, Murcia, Spain
        \vspace{-1mm} 
    \end{tabular}
    \caption{Datasets used in our experiments are collected using a GoPro-9 camera at three different locations. A sample for each dataset is shown; notice the (a) grey/white thinner cavelines in the \textit{Florida dataset}, the line starts near the lower right corner; (b) thick yellow cavelines and a decorated background in the \textit{Mexico dataset}, line starts near the lower left corner; and (c) thick orange cavelines in the \textit{Spain dataset}. }
    \label{fig:datasets}
\vspace{-3mm}
\end{figure*}
Furthermore, we developed a  post-processing algorithm to filter the raw output masks of CL-ViT for more consistent and smooth caveline localization. We achieve this by first extracting a set of candidate lines from the binary output mask, then applying a voting procedure for non-maxima suppression based on the \emph{accumulator space} of the probabilistic Hough transform~\cite{kiryati1991probabilistic}. We demonstrate the effectiveness of this post-processing step qualitatively for noisy predictions in various challenging scenarios as well. 

%achieves SOTA performance on standard benchmarks  
% cave formations (speleothems: stalactites, stalagmites, columns, \etc) together with navigational aids such as arrows and cookies.

%% file: src/Related_Work.tex
\section{Background \& Related Work}

\subsection{Underwater Cave Mapping and Exploration}
%\invis{Traditional maps of underwater caves are graphs where the edges consist of straight line segments of the caveline and vertices are the attachment points. The water depth is manually recorded for each attachment point together with the length  and the orientation (yaw) of each edge. Figure \ref{fig:Spain} presents such a map (black line) for the Cueva del Agua cave in Spain together with a trajectory estimated by VIO~\cite{JoshiICRA2022}. Recording all this information together with additional observations, such as distance to the walls, ceiling, and floor, is called surveying~\cite{Burge1988Survey}, and is a challenging, time\hyp consuming error\hyp prone process.}
In order to produce informative representations of underwater caves, divers typically use photogrammetry~\cite{fortin2021environmental}, with a special focus on recording archaeological sites~\cite{gonzalez2008arrival,rissolo2015novel}. Attempts to automate underwater cave mapping by AUVs have proven to be challenging and thus remain an open problem. Mallios \etal~\cite{mallios2016toward} deployed an AUV to manually collect acoustic data from inside a cave for offline mapping, whereas Weidner \etal~\cite{WeidnerICRA2017,WeidnerMSc2017} utilized a stereo camera to map the walls of a cave. It is worth noting that vision\hyp based underwater state estimation is extremely challenging due to the lighting variations, light absorption, and blurriness~\cite{JoshiIROS2019}. More recently, Rahman \etal~\cite{RahmanICRA2018,RahmanIROS2019a,RahmanIJRR2022} presented a framework where acoustic, visual, inertial, and water depth data are used to estimate the trajectory of the robot and also a sparse representation of the cave. Denser representation of the cave boundaries can be obtained by mapping the contours~\cite{massone2020contour}, the moving shadows~\cite{RahmanIROS2019b} or via dense stereo reconstruction~\cite{WangICRA2023}. The above-presented approaches will be utilized to enhance the mapping of an AUV following the caveline safely in and out of the cave. Sunfish~\cite{richmond2020autonomous} a new man\hyp portable AUV is currently being deployed in caves in Florida.

\subsection{Object Detection/Segmentation in Underwater Imagery}
An essential capability of visually guided AUVs is to identify relevant objects and interesting image regions to make effective navigational decisions in real time. Various model-based techniques are generally deployed in fast visual search~\cite{koreitem2020one,islam2022svam}, enhanced object detection~\cite{zhu2020saliency,islam2020fast}, and monitoring applications~\cite{ModasshirCRV2018,manderson2018vision}. For instance, Koreitem~\etal~\cite{koreitem2020one} used a bank of pre-specified image patches to learn a similarity operator that guides the robot's visual search in an unconstrained setting. Besides, model-free approaches are more feasible for autonomous exploratory applications~\cite{girdhar2016modeling}. For instance, Girdhar~\etal~\cite{girdhar2014autonomous} formulated an online topic-modeling scheme that encodes visible features into a low-dimensional semantic descriptor for AUV exploration. More recent work by Modasshir~\etal combined a deep learning-based classifier model with VIO to identify and track the locations of different types of corals to generate semantic maps~\cite{ModasshirRobio2018} as well as volumetric models~\cite{ModasshirFSR2019}.

Due to the difficulties in acquiring large-scale labeled underwater data, the existing systems attempt to collect and annotate small-scale application-specific image data~\cite{alonso2019coralseg,ravanbakhsh2015automated}.
Islam~\etal~\cite{islam2020suim} considered eight object categories for human-robot cooperative underwater missions: robots, human divers, wrecks/ruins, aquatic plants, fish, reefs, and sea floor. Other datasets consider even fewer object categories such as marine debris or ship hull defects~\cite{waszak2022semantic}. With limited training samples per object category over only a few waterbody types, it is extremely challenging to achieve good generalization performance by SOTA deep learning-based models for image recognition tasks. These limitations call for learning adaptations with limited supervision and rigorous model design to ensure robust underwater visual perception.

%% file: src/Methodology.tex
\section{Weakly Supervised Caveline Detection}
\begin{figure*}[t]
\centering    \includegraphics[width=\linewidth]{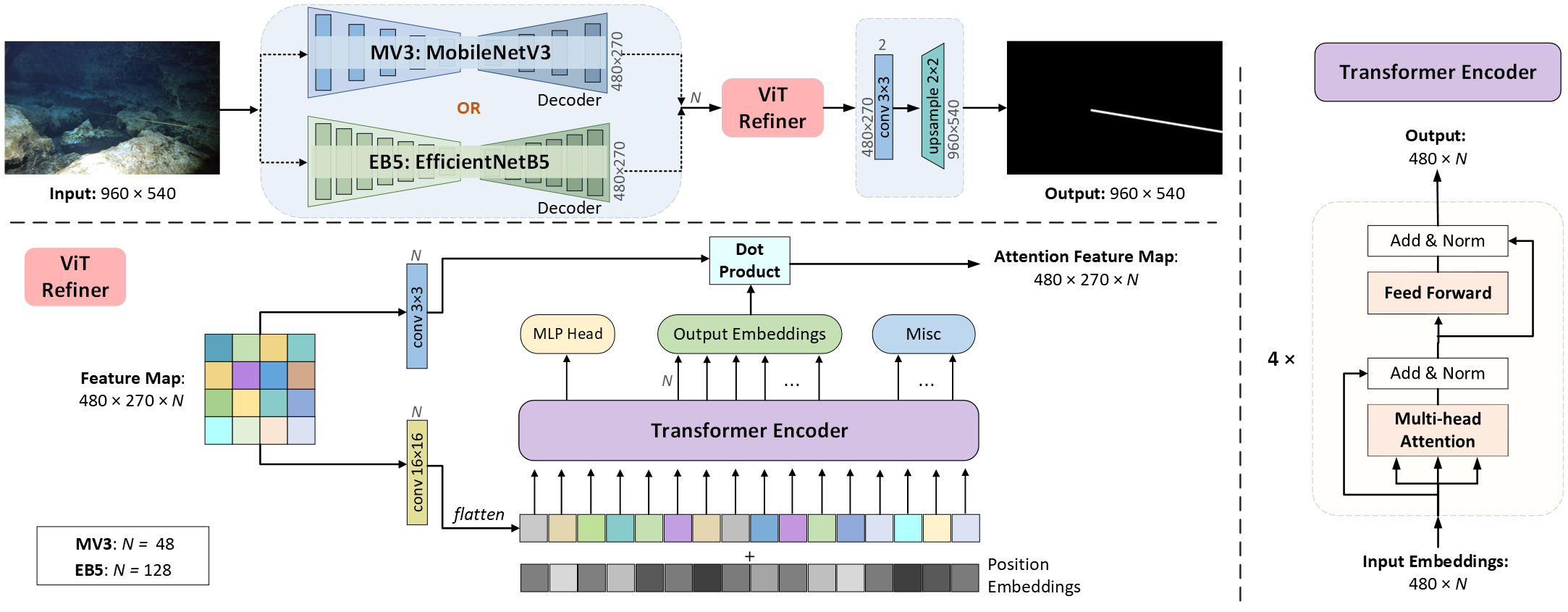}
\vspace{-1mm}
\caption{The end-to-end learning pipeline of our proposed CL-ViT model is shown. Input images are first fed to the backbone 
(light model: MobileNetV3 backbone; base model: EfficientNetB5 backbone) for feature extraction. Those features are forwarded to our transformer-based refinement module (ViT Refiner), followed by a convolution and upsampling block to generate the caveline detection mask. }
\label{fig:model_arch}
\vspace{-3mm}
\end{figure*}

\subsection{Problem Formulation and Data Preparation}
% list the problem
We formulate the problem of caveline detection in the RGB space as a binary image segmentation task, \ie, identifying pixels with caveline as a semantic map~\cite{hafiz2020survey}. In our task, the background pixels and caveline pixels are assigned with $0$ and $1$ labels, respectively. 
For data-driven training and evaluation, we extract video frames from our cave exploration experiments~\cite{JoshiICRA2022,RahmanIJRR2022} conducted in three different locations: the Devil's system in Florida, USA; the Dos Ojos Cenote, QR, Mexico; and the Cueva del Agua in Murcia, Spain. We grouped the caveline frames from these locations into three datasets, which we term as the \textit{Florida}, \textit{Mexico}, and \textit{Spain} dataset, respectively. 

As illustrated in Fig.~\ref{fig:datasets}, we found that the three cave locations exhibit different caveline characteristics in terms of thickness, color, and background patterns. The cavelines in Florida are thin and off\hyp white colored, whereas the Mexico caves are the most decorated with yellow colored lines. Cavelines in Spain are also thick and of orange color. In general, the main cavelines in popular locations are thicker, while off the main path become thinner; the grey/white colored lines take a darker color over time and often blend with the background patterns. We identify these variety of challenging cases and prepare $1050$ images in each set, totaling $3\times1050=3150$ instances. We focused on maximizing variance in the data by including varieties in caveline color, distance, background/waterbody patterns as well as different cave formations (\eg, stalactites, stalagmites, columns) and navigational aids such as arrows and cookies. Four human participants sorted these image samples and then pixel-annotated the cavelines for ground truth generation, which we utilizes for the training and evaluation of all models.

\subsection{Deep Visual Learning Framework}
We developed a unified training pipeline for SOTA semantic segmentation models across the CNN, CRF, and ViT literature. Specifically, we used the following models for the baseline performance analyses.
\begin{itemize}
    \vspace{-3mm}
    
    \item \textbf{UNet~\cite{ronneberger2015u}}: While originally proposed for medical image segmentation, many UNet variants~\cite{zhou2019unet++,zhang2018road,isensee2021nnu} have been proposed for different application domains over the years. UNet models employ an encoder-decoder design with mirrored skip-connections at each feature resolution within a hierarchical top-down architecture. They are known to achieve good binary image segmentation performance from limited training samples through data augmentation. %
    \vspace{-3mm}
    \item \textbf{EMANet~\cite{li2019expectation}}: It is based on the self-attention mechanism, and employs an Expectation-Maximization (EM) algorithm to find a compact basis to compute category-specific attention maps. It is known to be computationally efficient and robust to the variance of input through its low-rank factorization. %
    \vspace{-3mm}
    \item \textbf{Dense Prediction Transformer (DPT)~\cite{ranftl2021vision}}: It is a transformer-based architecture that assembles tokens from various stages of an encoder into image-like representations for multi-scale progressive learning. It employs a ViT~\cite{dosovitskiy2020image} backbone for dense prediction tasks; it is known for fine-grained and globally coherent semantic map generation.%
    \vspace{-3mm}
    \item \textbf{Pyramid Attention Network (PAN)~\cite{li2018pyramid}}: It integrates the {global attention} mechanism with a spatial pyramid to extract dense features for efficient semantic labeling. PAN-based learning pipelines with CNN-based encoder-decoders can learn to achieve good object localization performance and boundary details with low-resolution data~\cite{islam2020sesr,li2020scattnet}.  %
    \vspace{-3mm}
    \item \textbf{DeepLabv3+~\cite{chen2018encoder}}: It uses the notion of atrous~\cite{chen2017deeplab} spatial pyramid pooling (ASPP) to extract encoder features at arbitrary resolutions and then employs CRF-based post-processing stages to ensure multi-scale context awareness. DeepLab models~\cite{chen2018encoder,chen2017deeplab} are capable of achieving fine-grained semantic and instance segmentation performance across different scales and texture patterns. 
\end{itemize}

\subsection{Proposed Model: \textbf{CL-ViT}}
We develop a lightweight caveline detection model CL-ViT for use by visually guided AUVs in underwater cave mapping and exploration tasks. To this end, we focus on enabling two important features: $(i)$ robustness to noisy low-resolution inputs because cavelines are only a few pixels wide even in a high-resolution camera feed; and $(ii)$ efficient inference on single-board embedded platforms. We attempt to achieve this in CL-ViT model by integrating multi-scale local hierarchical features and global spatial information for efficient pixel-wise segmentation of cavelines. CL-ViT consists of two major learning components: an efficient encoder-decoder backbone and a ViT-based refinement module; the network architecture is illustrated in Fig.~\ref{fig:model_arch}. 

\subsubsection{Choice of Backbones}
We incorporate two options for the deep hierarchical feature extraction in CL-ViT: $(i)$ a light model with {MobileNetV3}~\cite{howard2019searching} backbone for on-board AUV processing; and $(ii)$ a base model with {EfficientNetB5~\cite{tan2019efficientnet}} backbone for offline use, \eg, when human operators on surface control ROVs inside a cave. The MobileNetV3 is a lightweight CNN-based model designed for resource-constrained platforms. The encoder contains a series of fully convolutional layers with $16$ filters followed by $15$ residual bottleneck layers. We then use a mirrored decoder with six convolutional blocks to map the encoded features into $48$ filters of $480 \times 270$ resolution (with an input of $960\times540\times3$). On the other hand, EfficientNet uses a technique called \emph{compound coefficient} to scale up models in a simple but effective manner. It uniformly scales features in width, depth, and resolution to ensure effective receptive fields for feature extraction. We use EfficientNetB5 which extracts $128$ filters of $480 \times 270$ resolution from $960 \times 540 \times 3$ inputs.   
%We apply the MobileNetV3 model~\cite{howard2019searching} as the backbone for pixel-level feature extraction due to its ability that maintains high accuracy while reducing the model size and computational complexity. MobileNetV3 implements a lightweight convolutional neural network for mobile devices by adopting optimization strategies such as fine-grained grid search algorithms and dynamic edge computing techniques, built on inverted residual blocks and linear bottleneck blocks. The encoder contains a series of fully convolution layers with $16$ filters using the hard swish nonlinearity followed by a total of $15$ residual bottleneck layers. The basic blocks of the convolutional layers form the decoder, where the $2\times$ bilinear upsampling of the previous block is concatenated with an encoder block with the same spatial size, resulting in a final $48$ filter generation of $480 \times 270$ resolution.

\subsubsection{ViT-based Refinement Module}
Following feature extraction, we design a ViT-based refinement module to \emph{transform} and \emph{embed} the contextual features into an efficient prediction head. Our idea is to allow each feature position to have consistent receptive fields so that the global spatial information is accurately embedded. As shown in Fig.~\ref{fig:model_arch}, the $N$=$48$/$128$ filters extracted by the backbone are $16 \times 16$ convolved and flattened to \emph{patch embeddings}, which are concatenated with learnable \emph{position embeddings}. These embeddings are then propagated to the transformer encoder, which applies a four-layer multi-head attention mapping. Subsequently, the normalized and $3\times3$ convolved feature maps are projected (dot product operation) with $N$ output embeddings; the remaining embeddings in the MLP head are dropped. The selected attention maps then generate the binary caveline segmentation map after a final convolution and upsampling operation at $960\times540$ resolution.

\subsubsection{Learning Objective}
The end-to-end training is driven by two loss functions: the standard cross-entropy loss~\cite{zhang2018generalized} and the Dice loss proposed by Milletari~\etal~\cite{milletari2016v}. The cross-entropy loss quantifies the dissimilarity in pixel intensity distributions between the generated caveline map ($\hat{y}$) and its ground truth ($y$). For a total of $n_p$ pixels, it is calculated as:
\begin{equation}
\footnotesize
    \mathcal{L}_{BCE} = \frac{1}{n_p} \sum\nolimits_{i} [-{y}_{i} \log \hat{y}_{i} - (1- {y}_{i})\log(1- \hat{y}_{i})].
\end{equation}
While we initially trained all caveline detectors with $\mathcal{L}_{BCE}$ alone, we noticed a severe class imbalance problem, since there are very few positive (caveline) pixels compared to the negative (background) pixels. We address this issue by adding the Dice loss, which balances foreground and background classes by normalizing ${y}$ and $\hat{y}$ as follows:
\begin{equation}
\footnotesize
\mathcal{L}_{Dice}=1- \frac{2\sum\nolimits_{i}{y}_{i}\hat{y}_{i}}{\sum\nolimits_{i}{y}_{i}^{2}+\sum\nolimits_{i}\hat{y}_{i}^{2}}.
\end{equation}
Finally, the end-to-end learning objective is formulated as:  
\begin{equation}
\footnotesize
 \mathcal{L}_{CL} = \lambda_{CE} \; \mathcal{L}_{BCE} + \lambda_{D}  \;\mathcal{L}_{Dice}.
\label{obj_fun}
\end{equation}
Here, we find the $\lambda_{CE}$ and $\lambda_D$ empirically for different models independently through hyper-parameters tuning.

\subsubsection{Weakly-Supervised Iterative Training}\label{weak_train}
\begin{figure}[h]
\centering    \includegraphics[width=\linewidth]{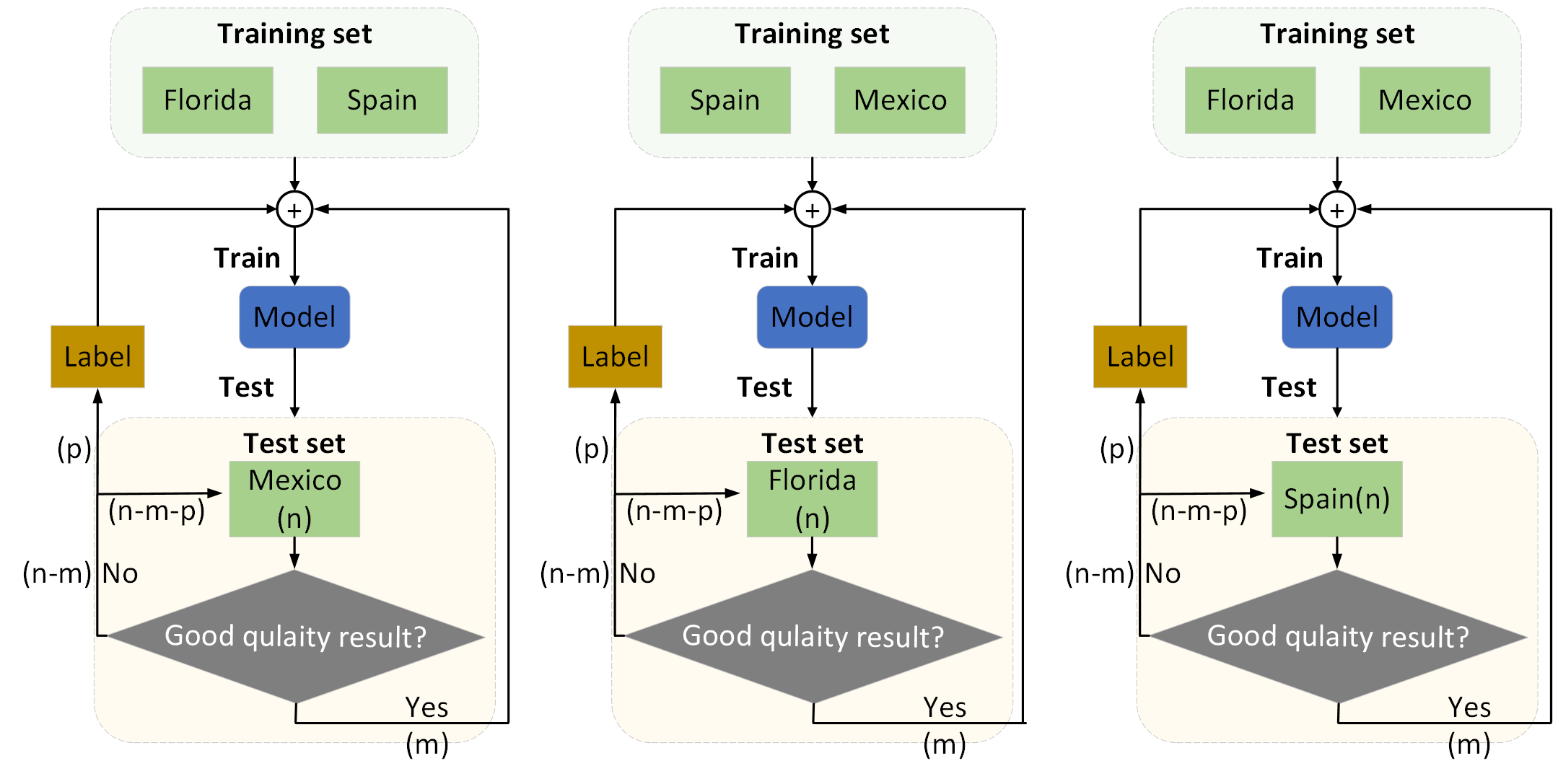}
\caption{Our weakly-supervised iterative training process is shown.}
\label{fig:iterative_train}
\vspace{-2mm}
\end{figure}

\begin{table*}[h]
\centering
\caption{Quantitative comparison for caveline detection performance by CL-ViT and other SOTA models are shown for our weakly supervised iterative learning phases (see Sec.~\ref{weak_train} for the discussion on how these training phases are carried out).}
\vspace{-1mm}
\scalebox{0.9}{
\begin{tabular}{l|l|c|c||c|c|c|c|c|c|c}
  \Xhline{2\arrayrulewidth}
   \hline
    \textbf{Training data} & \textbf{Test set} & \textbf{Phase} & \textbf{Metric ($\uparrow$)}  & EMANet & UNet & DPT & PAN & DeepLabv3+ & \makecell[c]{\textbf{CL-ViT} \\(MV3)} & \makecell[c]{\textbf{CL-ViT} \\(EB5)} \\
    
    \hline
    \multirow{5}*{Florida + Spain} & \multirow{5}*{Mexico} & $1$ & \makecell[l]{IoU \\ F1} &  \makecell[c]{$13.43$ \\$77.39$ } & \makecell[c]{ $32.65$\\ $66.35$} & \makecell[c]{$35.07$ \\$68.42$ } & \makecell[c]{ $38.39$ \\  $82.12$ } & \makecell[c]{ $46.16$ \\  $84.65$ } & \makecell[c]{ $25.08$\\ $59.04$}& \makecell[c]{ $\mathbf{50.92}$\\  $\mathbf{85.72}$ } \\
    
    \cline{3-11}
      &  & $2$ & \makecell[l]{IoU \\ F1} &  \makecell[c]{$15.90$ \\$83.35$ } & \makecell[c]{ $45.56$\\$82.50$ } & \makecell[c]{ $51.21$\\$89.29$ } & \makecell[c]{  $62.41$\\  $96.62$ } & \makecell[c]{  $61.90$\\  $95.99$ } & \makecell[c]{ $33.08$\\$73.95$ } & \makecell[c]{  $\mathbf{68.73}$\\  $\mathbf{97.84}$} \\

    \cline{3-11}
       & & $3$ & \makecell[l]{IoU \\ F1} &  \makecell[c]{$15.98$ \\$79.74$ } & \makecell[c]{$46.05$ \\ $83.45$} & \makecell[c]{$54.07$ \\$88.53$ } & \makecell[c]{ $55.84$ \\  $95.04$} & \makecell[c]{  $60.57$\\ $96.98$ } & \makecell[c]{$32.60$ \\$72.45$ } & \makecell[c]{  $\mathbf{70.28}$\\  $\mathbf{98.34}$} \\

       \hline
    \multirow{5}*{Florida + Mexico} & \multirow{5}*{ Spain} & $1$ & \makecell[l]{IoU \\ F1} &  \makecell[c]{$13.10$ \\$77.67$ } & \makecell[c]{  $55.32$\\$91.24$ } & \makecell[c]{$48.23$ \\$84.60$ } & \makecell[c]{$54.86$ \\  $93.65$} & \makecell[c]{  $58.19$\\  $95.43$} & \makecell[c]{$42.07$ \\$84.24$ } & \makecell[c]{  $\mathbf{66.47}$\\  $\mathbf{96.03}$} \\
    
    \cline{3-11}
      &  & $2$ & \makecell[l]{IoU \\ F1} &  \makecell[c]{ $24.45$\\$92.54$ } & \makecell[c]{ $60.30$\\$96.80$ } & \makecell[c]{$64.23$ \\$97.14$ } & \makecell[c]{ $66.78$ \\  $98.07$ } & \makecell[c]{  $70.70$\\  $98.47$ } & \makecell[c]{ $48.78$\\$92.92$ }  & \makecell[c]{  $\mathbf{76.00}$\\  $\mathbf{98.48}$}\\

    \cline{3-11}
       & & $3$ & \makecell[l]{IoU \\ F1} &  \makecell[c]{ $18.67$\\$79.13$ } & \makecell[c]{$62.88$ \\$96.93$ } & \makecell[c]{ $67.24$\\$97.06$ } & \makecell[c]{ $69.86$ \\  $98.39$ } & \makecell[c]{  $71.13$\\  $98.64$ } & \makecell[c]{$48.29$ \\$91.71$ }  & \makecell[c]{  $\mathbf{77.39}$\\  $\mathbf{98.74}$} \\

       \hline
    \multirow{5}*{Spain + Mexico} & \multirow{5}*{Florida} & $1$ & \makecell[l]{IoU \\ F1} &  \makecell[c]{\hspace{0.5mm} $6.87$ \\$31.45$ } & \makecell[c]{$19.21$ \\$37.50$ } & \makecell[c]{$18.89$ \\$35.45$ } & \makecell[c]{  $27.87$\\  $55.90$ } & \makecell[c]{  $28.78$\\  $56.50$ } & \makecell[c]{$16.18$ \\$37.68$ }  & \makecell[c]{  $\mathbf{39.11}$\\  $\mathbf{74.23}$} \\
    
    \cline{3-11}
      &  & $2$ & \makecell[l]{IoU \\ F1} &  \makecell[c]{ $13.60$\\$65.66$ } & \makecell[c]{ $24.24$\\$52.75$ } & \makecell[c]{ $26.17$\\$57.24$ } & \makecell[c]{ $28.53$ \\  $71.64$ } & \makecell[c]{  $33.30$\\  $77.73$ } & \makecell[c]{$20.64$ \\$54.49$ }  & \makecell[c]{  $\mathbf{41.16}$\\  $\mathbf{85.26}$} \\

    \cline{3-11}
       & & $3$ & \makecell[l]{IoU \\ F1} &  \makecell[c]{$13.20$ \\ $69.33$} & \makecell[c]{$26.49$ \\ $57.79$} & \makecell[c]{$22.01$ \\$51.23$ } & \makecell[c]{ $34.81$ \\  $79.21$ } & \makecell[c]{  $31.67$\\  $76.71$ } & \makecell[c]{$15.79$ \\$47.69$ } & \makecell[c]{  $\mathbf{40.53}$\\  $\mathbf{85.00}$} \\

    \Xhline{2\arrayrulewidth} 
\end{tabular}}
\label{tab_com_datasets}
%\vspace{-3mm}
\end{table*}

\begin{figure*}[t]
\centering
\includegraphics[width=0.79\linewidth]{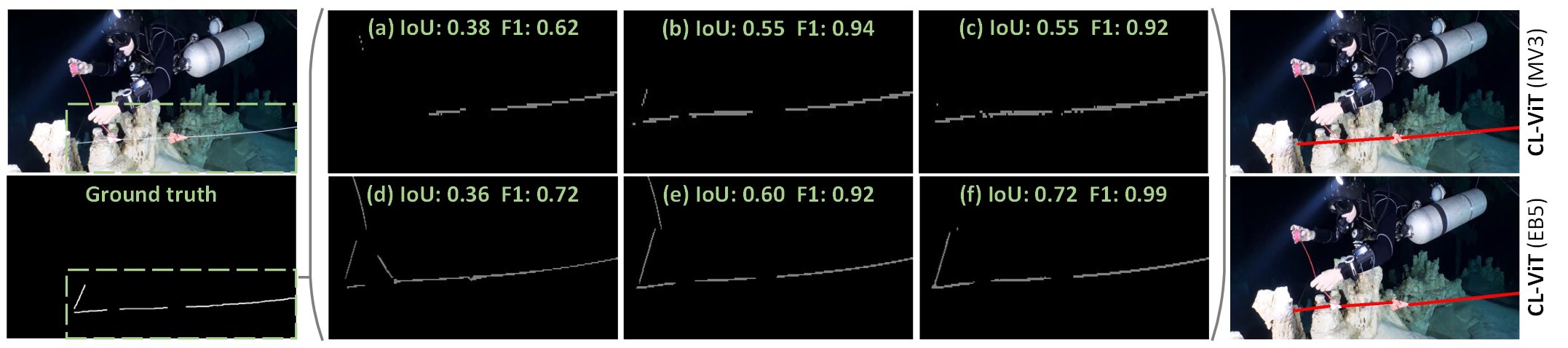}~
\includegraphics[width=0.19\linewidth]{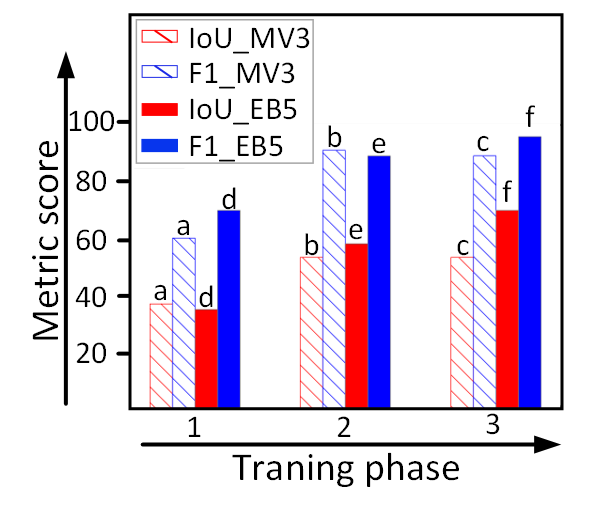}%
\vspace{-3mm}
\caption{Effectiveness of our weakly supervised caveline detection pipeline is shown with an example. Output maps a/d, b/e, and c/f are the test results for CL-ViT model with MobileNetV3/EfficientNetB5 backbone after the first, second, and third phase of training, respectively. As seen, the visual prediction results and metric scores gradually get better after each learning phase. The final post-processed predictions are also shown on the right overlayed on the input image.}
\label{fig:iterative_comp}
\vspace{-3mm}
\end{figure*}

Pixel-annotated training data is very scarce for unique problems such as caveline detection in underwater caves. As discussed earlier, caveline characteristics and background waterbody patterns in each cave locations differ greatly, making it difficult to compile a comprehensive dataset for supervised training. We address this limitation by a weakly supervised formulation, where model accuracy is improved incrementally on new locations' data. This speeds up model adaptations during robotics field deployments to a new location by eliminating the need for labeling entire datasets for supervised training. We validate this hypothesis on each dataset (\ie, Florida, Spain, and Mexico data) based on the \textit{leave-one-out} mechanism, as illustrated in Fig.~\ref{fig:iterative_train}. 

For each of the three cases shown in Fig.~\ref{fig:iterative_train}, the weak supervision is carried out as follows. The initial model is evaluated on the full test set (of $1050$ samples), from which a human expert sorts out the good quality predictions to reinforce the learning in the next phase. The human expert also selects a set of challenging samples where the model failed, then annotates and combines them into the training set in order to balance distribution positive (accurate) and negative (erroneous) samples. This process is repeated several times until a satisfactory number of training samples are compiled. Our experiments reveal that we get $15\%$-$20\%$ good quality predictions in the first phase and another $34\%$-$47\%$ in the second phase. All $1050$ images get labelled within $3$ phases, where human experts relabelled only $200$-$250$ images as negative samples. Thus the remaining labels are \textit{weak labels} generated by intermediate sub-optimal models for subsequent weak supervision.   
\subsubsection{Post\hyp processing}
In the post-processing step, we smooth the raw CL-ViT output of binary pixel predictions into continuous line segments. We achieve this by first interpolating a sequence of connected straight line segments by modeling a probabilistic Hough transform~\cite{kiryati1991probabilistic}. Then we apply a voting procedure for non-maxima suppression and generate the most dominant line. To ensure robustness of this suppression mechanism for all types of noisy and incomplete predictions, we empirically tuned the hyper-parameters, \eg, the distance metric, acute angle threshold for merging pair-wise lines, and the number of iterations. 
%In a post-processing step, we were able to extract analytical description of the cavelines from segmentation mask generated by the neural network. As deployed cavelines appear as fairly straight line segments in most typical scenarios, 
%One challenge of the process was suppression of redundant line-segments which are typical in output of Hough transform. We tried morphological techniques like erosion on the input mask without much success, as they tend to introduce discontinuities. Our working solution uses an iterative approach to merge line segment pairs who make an acute angle less than some specified angle and also have the shortest euclidean distance less than some other specified value.

\begin{figure*}[t]
\centering
\includegraphics[width=\linewidth]{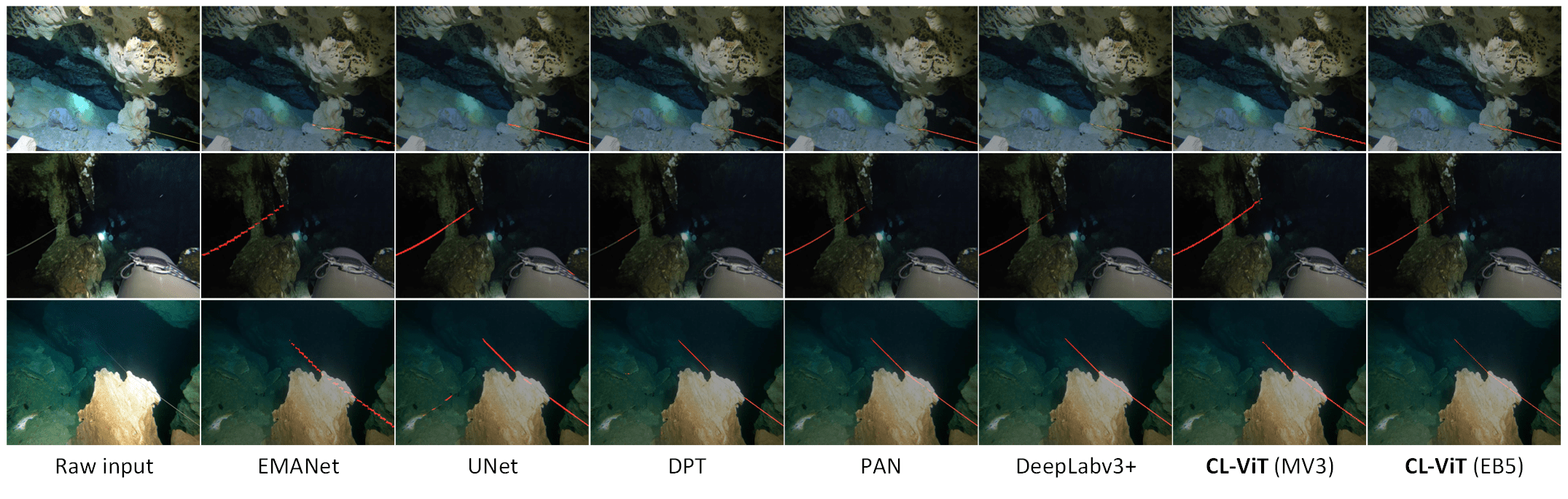}
\vspace{-7mm}
\caption{A few qualitative comparisons are shown for caveline detection by CL-ViT and other SOTA models on cross-location test images. The DeepLabv3+, DPT, and PAN models provide well-localized predictions, while CL-ViT (EfficientNetB5) generates the most fine-grained caveline detection (\ie, thinnest continuous lines). Note that all images are overlayed with raw outputs without post-processing. 
%As seen, our lightweight MobileNetV3-based CL-ViT offers comparable visual guide performance with its competitors while offering $10 \times $ faster inference and memory efficiency for on-board AUV use, and our EfficientNet-based CL-ViT provides the most accurate caveline detection results for off-line use. 
}
\label{fig:qual_comp}
\vspace{-5mm}
\end{figure*}

%% file: src/Experiments.tex
\section{Experimental Results}
\subsection{Baseline Models and Evaluation metrics}

We developed a unified training pipeline for SOTA models across the CNN, CRF, and ViT literature. Specifically, we use:    EMANet~\cite{li2019expectation}, 
UNet~\cite{ronneberger2015u},
DPT~\cite{ranftl2021vision}.
PAN~\cite{li2018pyramid}, and
DeepLabv3+~\cite{chen2018encoder} for baseline performance analyses. We use Pytorch libraries to implement a unified learning pipelines for CL-ViT and all SOTA models in comparison. RMSprop~\cite{tieleman2012lecture} is used as the optimizer with an initial learning rate of $10^{-5}$, a momentum of $0.9$, and a weight decay of $10^{-8}$. The input-output resolution is set to $960\times540$ for all models; other SOTA model-specific parameters are chosen based on their respective recommended configurations.

For performance evaluation, we use two standard metrics: \textbf{IOU} and \textbf{F1 score}. The {IoU} ({Intersection Over Union)} measures caveline localization performance using the area of overlapping regions of the predicted and ground truth labels. it is defined as $IoU = \frac{Area\text{ }of\text{ }overlap}{Area\text{ }of\text{ }union}$. 
Besides, the {F1 score} quantifies the correctness of predicted labels compared to ground truth by the normalized precision ($\mathcal{P}$) and recall ($\mathcal{R}$) scores as $\mathcal{F}=\frac{2 \times \mathcal{P} \times \mathcal{R}}{\mathcal{P}+\mathcal{R}}$.

%The proposed and comparable models are supervised by RGB images of underwater caves containing cavelines paired with binary segmentation labels. $95\%$ of images in the training dataset are used for training, and the remaining $5\%$ are used for validation. 

%%For specific model structure applications, Resnet-50~\cite{he2016deep} is used as an encoder with the pre-trained weight of the ``Semi-weakly" supervised (SWSL) ImageNet model~\cite{yalniz2019billion} for training the DeepLabv3+~\cite{chen2018encoder} and the PAN~\cite{li2018pyramid}. EMANet~\cite{li2019expectation} was trained with the backbone ResnNet-101~\cite{he2016deep} as its excellent performance in other segmentation tasks. Considering the limited cost platform required for the caveline detection task, we adopt the relatively lightweight Hybrid version for the DPT~\cite{ranftl2021vision} model, which uses the ViT-Hybrid that consists of a ResNet-50 followed by 12 transformer layers. Our proposed CL-ViT model adopts the MobileNetV3-Large and EfficientNet-B5 as the two options for encoder with no additional modifications.

\subsection{Qualitative and Quantitative Evaluation}
\subsubsection{Effectiveness of Weak Supervision}
We first demonstrate the utility and effectiveness of our weakly supervised learning pipeline for the three cases depicted in 
 Fig.~\ref{fig:iterative_train}. The corresponding quantitative results are listed in Table~\ref{tab_com_datasets}, which shows that all models exhibit incremental improvements over learning phases $1$, $2$, and $3$. This validates our intuition that robust generalization performance by standard deep visual learning models can be achieved with very few labeled data from a new location for fast model adaptation. Fig.~\ref{fig:iterative_comp} shows a particular example where IoU scores improved from $0.36$/$0.38$ to $0.55$/$0.72$, while F1 scores improved from $0.62$/$0.72$ to $0.92$/$0.99$ for the CL-ViT light/base model, respectively. The generated maps become increasingly fine-grained as well; the final output maps can be further post-processed for well-localized detection of cavelines.    

%: With the training iterations, the detected cavelines are gradually approaching the groundtruth, which is also proved by the scores; and the final post-processing results provide more straightforward visual results than those recognized by human eyes, enabling the caveline detection and tracking ability for AUVs.  

%problem with only manually labeling $20\%$ to $30\%$ of the training data. In the second phase, the IoU metric of all models reveal an average improvement of $27.09\%$ on three different data sets, and the F1 metric increased by $22.66\%$. Although the metric value of the third phase is not increased much, the qualitative results provide a more stable prediction: less noise, smoother and more complete cavelines. 

\subsubsection{Performance Analyses of CL-ViT}
We conduct a thorough performance evaluation of CL-ViT and other SOTA models based on all cross-location test images. A few qualitative comparisons are shown in Fig.~\ref{fig:qual_comp}, which shows consistent results from CL-ViT, DeepLabv3+, DPT, and PAN. Our CL-ViT (EfficentNetB5) model achieves the most fine-grained caveline detection performance with the thinnest continuous line segments. While not as fine-grained, oue light CL-ViT (MobileNetV3) model localizes the cavelines reasonably as well, which can be further refined by post-processing. A video demonstration can be seen here: \url{https://youtu.be/AXYHlaAw-Ig}.

\begin{table}[h]
\centering
\caption{Quantitative test results are shown for CL-ViT and other \textbf{top three} models from Table~\ref{tab_com_datasets}; their memory requirements in Mega-Bytes (MB), and inference rates in FPS (Core i9-12900 CPU) and FPS$^*$ (single-board Jetson TX2) are compared as well.}
\vspace{-1mm}
\scalebox{0.85}{
\begin{tabular}{l||c|c|c|c|c}
  \Xhline{2\arrayrulewidth}
   \hline \textbf{Metric} & UNet & DPT &  DeepLabv3+ & \makecell[c]{\textbf{CL-ViT} \\(MV3)} & \makecell[c]{\textbf{CL-ViT} \\(EB5)} \\ 
    \Xhline{2\arrayrulewidth} 
    $\uparrow$ IoU & $38.34$ & $38.88$ & $49.77$ & $28.57$ & $\mathbf{58.30}$ \\
    $\uparrow$ F1 & $86.68$ & $77.89$ & $93.07$ & $77.79$ &  $\mathbf{95.87}$ \\
    \Xhline{2\arrayrulewidth} 
    $\uparrow$ FPS & $2.34$ & $0.46$ & $2.41$ & $\mathbf{20.21}$ & $0.77$  \\
    $\uparrow$ FPS$^{*}$ & $1.15$ & $0.23$ & $1.19$ & $\mathbf{10.71}$ & $0.38$  \\ 
    $\downarrow$ MB & $124.20$ & $496.20$ & $105.00$ & $\mathbf{50.90}$ & $313.70$ \\
    \Xhline{2\arrayrulewidth} 
\end{tabular}}
\label{tab_com_detectors}
\vspace{-2mm}
\end{table}

%The qualitative results corroborate our analyses; as the visual comparisons of Fig.~\ref{fig:qual_comp} illustrate, the EfficientNet-based CL-ViT provides the most accurate results for caveline detection (other models produce seemingly more visible effects due to pixel prediction deviations, while our model precisely detects tiny caveline pixels), and MobileNetV3-based CL-ViT achieves comparable or even better performance than SOTA models.

We show the corresponding quantitative test results in Table~\ref{tab_com_detectors}; it confirms the superior performance from CL-ViT (EfficentNetB5) for both IoU and F1 score metrics. Although CL-ViT (MobileNetV3) does not surpass the SOTA performance, with a significantly lighter model architecture, it offers $51.52\%$-$89.74\%$ memory efficiency and $7$-$43$ times faster inference rates. It runs at \textbf{$\mathbf{10}$+ FPS rates on Nvidia\texttrademark~Jetson TX2} devices, which makes it feasible for single-board deployments in AUVs' autonomy pipeline. 

\subsubsection{Challenging Cases}
\begin{figure}[t]
\centering
\includegraphics[width=\linewidth]{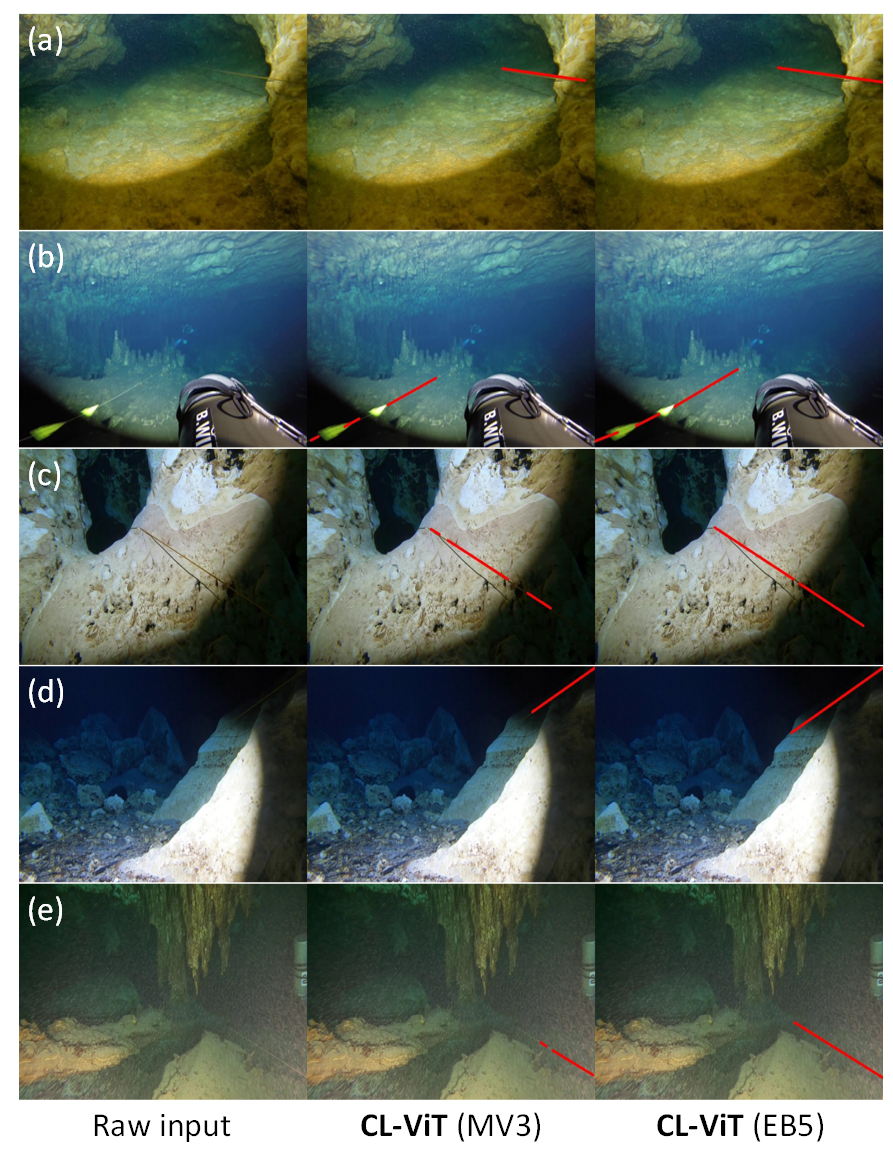}
\vspace{-7mm}
\caption{Post-processed CL-ViT output for a few challenging cases from the \textbf{CL-Challenge} test set are shown; notice the (a) lack of contrast/color difference between the caveline and background; (b) presence of arrows/cookies; (c) caveline shadow appears similar to another line; (d) caveline is outside the illuminated area; (e) scattering and optical distortions around the caveline.}
\label{fig:challenge}
\vspace{-6mm}
\end{figure}

As discussed earlier, very low resolutions of positive (caveline) pixels compared to the negative (background) pixels cause the \textit{class imbalance} problem in caveline detection learning. While we eliminate this by using a relatively high input resolution of $960\times540$, there are other challenging scenarios where CL-ViT (and other models) are faced with challenges. We identify a subset of such challenging cases and compile a \textbf{CL-Challenge} test set with $200$ samples. It includes images with severe optical distortions, lack of contrast, over-saturation, shadows, low-light conditions, occlusion, and other issues that make it extremely challenging to locate/segment the caveline, even for a human observer. As shown in Fig.~\ref{fig:challenge}, CL-ViT models are still able to localize the caveline for the most part. Despite some noisy predictions, these inspiring results indicate that caveline detection by CL-ViT can facilitate safe AUV navigation inside underwater caves.               

%% file: src/Conclusion.tex
\section{Conclusions}
In this paper, we presented a novel learning pipeline for fast caveline detection in images from underwater caves. We formulated a weakly supervised approach that facilitates a rapid model adaptation to data from new location by requiring very few ground truth labels. A comparison with SOTA frameworks demonstrated higher accuracy and efficiency of the proposed approach. Cavelines traverse the majority of explored underwater caves providing a roadmap that can guide a robot inside a cave and then safely back out. Of paramount importance is robustness across different appearances both of the line but also of the surrounding background. Tests on three different locales demonstrated accurate performance across different domains and lines. Currently, we are also investigating the automatic labeling of different cave formations (\eg, speleothems: stalactites, stalagmites, columns) together with navigational aids such as arrows and cookies. Our immediate next step is to develop an autonomous caveline\hyp following system that exploits CL-ViT's caveline predictions in tandem with a VIO system for visual servoing. Such autonomous operations inside caves will potentially lead to high-definition photorealistic map generation and more accurate volumetric models. 

%% file: main.bbl
\begin{thebibliography}{10}\itemsep=-1pt

\bibitem{alonso2019coralseg}
I.~Alonso, M.~Yuval, G.~Eyal, T.~Treibitz, and A.~C. Murillo.
\newblock {CoralSeg: Learning Coral Segmentation from Sparse Annotations}.
\newblock {\em {Journal of Field Robotics (JFR)}}, 36(8):1456--1477, 2019.

\bibitem{Burge1988Survey}
J.~Burge.
\newblock {\em Underwater Cave Surveying}.
\newblock Cave Diving Section of the National Speleological Society, 1988.

\bibitem{buzzacott2009american}
P.~L. Buzzacott, E.~Zeigler, P.~Denoble, and R.~Vann.
\newblock American cave diving fatalities 1969-2007.
\newblock {\em International Journal of Aquatic Research and Education},
  3(2):7, 2009.

\bibitem{chen2017deeplab}
L.-C. Chen, G.~Papandreou, I.~Kokkinos, K.~Murphy, and A.~L. Yuille.
\newblock {DeepLab: Semantic Image Segmentation with Deep Convolutional Nets,
  Atrous Convolution, and Fully Connected CRFs}.
\newblock {\em {IEEE Transactions on Pattern Analysis and Machine
  Intelligence}}, 40(4):834--848, 2017.

\bibitem{chen2018encoder}
L.-C. Chen, Y.~Zhu, G.~Papandreou, F.~Schroff, and H.~Adam.
\newblock Encoder-decoder with atrous separable convolution for semantic image
  segmentation.
\newblock In {\em Proceedings of the European conference on computer vision
  (ECCV)}, pages 801--818, 2018.

\bibitem{dosovitskiy2020image}
A.~Dosovitskiy, L.~Beyer, A.~Kolesnikov, D.~Weissenborn, X.~Zhai,
  T.~Unterthiner, M.~Dehghani, M.~Minderer, G.~Heigold, S.~Gelly, et~al.
\newblock An image is worth 16x16 words: Transformers for image recognition at
  scale.
\newblock {\em arXiv preprint arXiv:2010.11929}, 2020.

\bibitem{exley1986basic}
S.~Exley.
\newblock {\em Basic cave diving: A blueprint for survival}.
\newblock Cave Diving Section of the National Speleological Society, 1986.

\bibitem{karstbook}
D.~Ford and P.~Williams.
\newblock {\em Introduction to Karst}, chapter~1, pages 1--8.
\newblock John Wiley \& Sons, Ltd, 2007.

\bibitem{fortin2021environmental}
J.~Fortin, S.~Meacham, D.~Rissolo, C.~Le~Maillot, and F.~Devos.
\newblock Environmental challenges, technical solutions and standard operating
  procedures for data collection in photogrammetric studies toward a unified
  database of objects and features in underwater caves in mexico.
\newblock {\em The International Archives of Photogrammetry, Remote Sensing and
  Spatial Information Sciences}, 43:659--666, 2021.

\bibitem{girdhar2016modeling}
Y.~Girdhar and G.~Dudek.
\newblock {Modeling Curiosity in a Mobile Robot for Long-term Autonomous
  Exploration and Monitoring}.
\newblock {\em Autonomous Robots}, 40(7):1267--1278, 2016.

\bibitem{girdhar2014autonomous}
Y.~Girdhar, P.~Giguere, and G.~Dudek.
\newblock {Autonomous Adaptive Exploration using Realtime Online Spatiotemporal
  Topic Modeling}.
\newblock {\em {International Journal of Robotics Research (IJRR)}},
  33(4):645--657, 2014.

\bibitem{gonzalez2008arrival}
A.~H.~G. Gonz{\'a}lez, C.~R. Sandoval, A.~T. Mata, M.~B. Sanvicente, and
  E.~Acevez.
\newblock {The arrival of humans on the Yucatan Peninsula: Evidence from
  submerged caves in the state of Quintana Roo, Mexico}.
\newblock {\em Current Research in the Pleistocene}, 25:1--24, 2008.

\bibitem{hafiz2020survey}
A.~M. Hafiz and G.~M. Bhat.
\newblock A survey on instance segmentation: state of the art.
\newblock {\em International journal of multimedia information retrieval},
  9(3):171--189, 2020.

\bibitem{howard2019searching}
A.~Howard, M.~Sandler, G.~Chu, L.-C. Chen, B.~Chen, M.~Tan, W.~Wang, Y.~Zhu,
  R.~Pang, V.~Vasudevan, et~al.
\newblock Searching for mobilenetv3.
\newblock In {\em {IEEE/CVF International Conference on Computer Vision
  (ICCV)}}, pages 1314--1324, 2019.

\bibitem{isensee2021nnu}
F.~Isensee, P.~F. Jaeger, S.~A. Kohl, J.~Petersen, and K.~H. Maier-Hein.
\newblock nnu-net: a self-configuring method for deep learning-based biomedical
  image segmentation.
\newblock {\em Nature methods}, 18(2):203--211, 2021.

\bibitem{islam2020suim}
M.~J. Islam, C.~Edge, Y.~Xiao, P.~Luo, M.~Mehtaz, C.~Morse, S.~S. Enan, and
  J.~Sattar.
\newblock {Semantic Segmentation of Underwater Imagery: Dataset and Benchmark}.
\newblock In {\em {IEEE/RSJ International Conference on Intelligent Robots and
  Systems (IROS)}}. IEEE/RSJ, 2020.

\bibitem{islam2020sesr}
M.~J. Islam, P.~Luo, and J.~Sattar.
\newblock {Simultaneous Enhancement and Super-Resolution of Underwater Imagery
  for Improved Visual Perception}.
\newblock In {\em Robotics: Science and Systems (RSS)}, Corvalis, Oregon, USA,
  July 2020.

\bibitem{islam2022svam}
M.~J. Islam, R.~Wang, and J.~Sattar.
\newblock {SVAM: Saliency-guided Visual Attention Modeling by Autonomous
  Underwater Robots}.
\newblock In {\em Robotics: Science and Systems (RSS)}, NY, USA, 2022.

\bibitem{islam2020fast}
M.~J. Islam, Y.~Xia, and J.~Sattar.
\newblock {Fast Underwater Image Enhancement for Improved Visual Perception}.
\newblock {\em IEEE Robotics and Automation Letters (RA-L)}, 5(2):3227--3234,
  2020.

\bibitem{JoshiIROS2019}
B.~Joshi, S.~Rahman, M.~Kalaitzakis, B.~Cain, J.~Johnson, M.~Xanthidis,
  N.~Karapetyan, A.~Hernandez, A.~{Quattrini Li}, N.~Vitzilaios, and
  I.~Rekleitis.
\newblock {Experimental Comparison of Open Source Visual-Inertial-Based State
  Estimation Algorithms in the Underwater Domain}.
\newblock In {\em IEEE/RSJ International Conference on Intelligent Robots and
  Systems (IROS)}, pages 7221--7227, 2019.

\bibitem{JoshiICRA2022}
B.~Joshi, M.~Xanthidis, S.~Rahman, and I.~Rekleitis.
\newblock High definition, inexpensive, underwater mapping.
\newblock In {\em IEEE International Conference on Robotics and Automation
  (ICRA)}, pages 1113--1121, Philadelphia, PA, USA, 2022.

\bibitem{kiryati1991probabilistic}
N.~Kiryati, Y.~Eldar, and A.~M. Bruckstein.
\newblock A probabilistic hough transform.
\newblock {\em Pattern recognition}, 24(4):303--316, 1991.

\bibitem{koreitem2020one}
K.~Koreitem, F.~Shkurti, T.~Manderson, W.-D. Chang, J.~C.~G. Higuera, and
  G.~Dudek.
\newblock One-shot informed robotic visual search in the wild.
\newblock In {\em IEEE/RSJ International Conference on Intelligent Robots and
  Systems (IROS)}, pages 5800--5807. IEEE, 2020.

\bibitem{li2020scattnet}
H.~Li, K.~Qiu, L.~Chen, X.~Mei, L.~Hong, and C.~Tao.
\newblock Scattnet: Semantic segmentation network with spatial and channel
  attention mechanism for high-resolution remote sensing images.
\newblock {\em IEEE Geoscience and Remote Sensing Letters}, 18(5):905--909,
  2020.

\bibitem{li2018pyramid}
H.~Li, P.~Xiong, J.~An, and L.~Wang.
\newblock Pyramid attention network for semantic segmentation.
\newblock {\em arXiv preprint arXiv:1805.10180}, 2018.

\bibitem{li2019expectation}
X.~Li, Z.~Zhong, J.~Wu, Y.~Yang, Z.~Lin, and H.~Liu.
\newblock Expectation-maximization attention networks for semantic
  segmentation.
\newblock In {\em IEEE/CVF Int. Conference on Computer Vision}, pages
  9167--9176, 2019.

\bibitem{mallios2016toward}
A.~Mallios, P.~Ridao, D.~Ribas, M.~Carreras, and R.~Camilli.
\newblock Toward autonomous exploration in confined underwater environments.
\newblock {\em Journal of Field Robotics}, 33(7):994--1012, 2016.

\bibitem{manderson2018vision}
T.~Manderson, J.~C.~G. Higuera, R.~Cheng, and G.~Dudek.
\newblock {Vision-based Autonomous Underwater Swimming in Dense Coral for
  Combined Collision Avoidance and Target Selection}.
\newblock In {\em {IEEE/RSJ International Conference on Intelligent Robots and
  Systems (IROS)}}, pages 1885--1891. IEEE, 2018.

\bibitem{massone2020contour}
Q.~Massone, S.~Druon, Y.~Breux, and J.~Triboulet.
\newblock {Contour-based approach for 3D mapping of underwater galleries}.
\newblock In {\em Global Oceans 2020: Singapore--US Gulf Coast}, pages 1--6.
  IEEE, 2020.

\bibitem{milletari2016v}
F.~Milletari, N.~Navab, and S.-A. Ahmadi.
\newblock {V-net: Fully convolutional neural networks for volumetric medical
  image segmentation}.
\newblock In {\em {International Conference on 3D vision (3DV)}}, pages
  565--571. Ieee, 2016.

\bibitem{ModasshirCRV2018}
M.~Modasshir, A.~{Quattrini Li}, and I.~Rekleitis.
\newblock Deep neural networks: a comparison on different computing platforms.
\newblock In {\em Canadian Conference on Computer and Robot Vision (CRV)},
  pages 383--389, Toronto, ON, Canada, May 2018.

\bibitem{ModasshirFSR2019}
M.~Modasshir, S.~Rahman, and I.~Rekleitis.
\newblock {Autonomous 3D Semantic Mapping of Coral Reefs}.
\newblock In {\em 12th Conference on Field and Service Robotics (FSR)}, pages
  365--379, Tokyo, Japan, Aug. 2019.

\bibitem{ModasshirRobio2018}
M.~Modasshir, S.~Rahman, O.~Youngquist, and I.~Rekleitis.
\newblock {Coral Identification and Counting with an Autonomous Underwater
  Vehicle}.
\newblock In {\em IEEE International Conference on Robotics and Biomimetics
  (ROBIO)}, pages 524--529, Kuala Lumpur, Malaysia, (Finalist of T. J. Tarn
  Best Paper in Robotics), Dec. 2018.

\bibitem{ModasshirICRA2020}
M.~Modasshir and I.~Rekleitis.
\newblock Augmenting coral reef monitoring with an enhanced detection system.
\newblock In {\em IEEE International Conference on Robotics and Automation},
  pages 1874--1880, Paris, France, 2020.

\bibitem{RahmanICRA2018}
S.~Rahman, A.~{Quattrini Li}, and I.~Rekleitis.
\newblock {Sonar Visual Inertial SLAM of Underwater Structures}.
\newblock In {\em IEEE International Conference on Robotics and Automation},
  pages 5190--5196, 2018.

\bibitem{RahmanIROS2019a}
S.~Rahman, A.~{Quattrini Li}, and I.~Rekleitis.
\newblock {An Underwater SLAM System using Sonar, Visual, Inertial, and Depth
  Sensor}.
\newblock In {\em IEEE/RSJ International Conference on Intelligent Robots and
  Systems (IROS)}, pages 1861--1868, Macau, (IROS ICROS Best Application Paper
  Award. Finalist), 2019.

\bibitem{RahmanIROS2019b}
S.~Rahman, A.~{Quattrini Li}, and I.~Rekleitis.
\newblock Contour based reconstruction of underwater structures using sonar,
  visual, inertial, and depth sensor.
\newblock In {\em IEEE/RSJ International Conference on Intelligent Robots and
  Systems (IROS)}, pages 8048--8053, Macau, Nov. 2019.

\bibitem{RahmanIJRR2022}
S.~Rahman, A.~{Quattrini Li}, and I.~Rekleitis.
\newblock {SVIn2: A Multi-sensor Fusion-based Underwater SLAM System}.
\newblock {\em International Journal of Robotics Research},
  41(11-12):1022--1042, July 2022.

\bibitem{ranftl2021vision}
R.~Ranftl, A.~Bochkovskiy, and V.~Koltun.
\newblock Vision transformers for dense prediction.
\newblock In {\em IEEE/CVF Int. Conference on Computer Vision}, pages
  12179--12188, 2021.

\bibitem{ravanbakhsh2015automated}
M.~Ravanbakhsh, M.~R. Shortis, F.~Shafait, A.~Mian, E.~S. Harvey, and J.~W.
  Seager.
\newblock {Automated Fish Detection in Underwater Images Using Shape-Based
  Level Sets}.
\newblock {\em {Photogrammetric Record}}, 30(149):46--62, 2015.

\bibitem{richmond2020autonomous}
K.~Richmond, C.~Flesher, N.~Tanner, V.~Siegel, and W.~C. Stone.
\newblock {Autonomous exploration and 3-D mapping of underwater caves with the
  human-portable SUNFISH{\textregistered} AUV}.
\newblock In {\em Global Oceans 2020: Singapore--US Gulf Coast}, pages 1--10.
  IEEE, 2020.

\bibitem{rissolo2015novel}
D.~Rissolo, A.~N. Blank, V.~Petrovic, R.~C. Arce, C.~Jaskolski, P.~L.
  Erreguerena, and J.~C. Chatters.
\newblock {Novel application of 3D documentation techniques at a submerged Late
  Pleistocene cave site in Quintana Roo, Mexico}.
\newblock In {\em Digital Heritage}, volume~1, pages 181--182, 2015.

\bibitem{ronneberger2015u}
O.~Ronneberger, P.~Fischer, and T.~Brox.
\newblock U-net: Convolutional networks for biomedical image segmentation.
\newblock In {\em Medical Image Computing and Computer-Assisted Intervention
  (MICCAI)}, pages 234--241. Springer, 2015.

\bibitem{tan2019efficientnet}
M.~Tan and Q.~Le.
\newblock Efficientnet: Rethinking model scaling for convolutional neural
  networks.
\newblock In {\em International Conference on Machine Learning}, pages
  6105--6114. PMLR, 2019.

\bibitem{tieleman2012lecture}
T.~Tieleman, G.~Hinton, et~al.
\newblock Lecture 6.5-rmsprop: Divide the gradient by a running average of its
  recent magnitude.
\newblock {\em COURSERA: Neural networks for machine learning}, 4(2):26--31,
  2012.

\bibitem{WangICRA2023}
W.~Wang, B.~Joshi, N.~Burgdorfer, K.~Batsos, A.~{Quattrini Li}, P.~Mordohai,
  and I.~Rekleitis.
\newblock {Real-Time Dense 3D Mapping of Underwater Environments}.
\newblock In {\em IEEE International Conference on Robotics and Automation
  (ICRA)}, London, UK, 2023.

\bibitem{waszak2022semantic}
M.~Waszak, A.~Cardaillac, B.~Elves{\ae}ter, F.~R{\o}d{\o}len, and M.~Ludvigsen.
\newblock Semantic segmentation in underwater ship inspections: Benchmark and
  data set.
\newblock {\em IEEE Journal of Oceanic Engineering}, 2022.

\bibitem{WeidnerMSc2017}
N.~Weidner.
\newblock {Underwater Cave Mapping and Reconstruction Using Stereo Vision}.
\newblock Master's thesis, Computer Science and Engineering Department,
  University of South Carolina, Columbia, SC, 2017.

\bibitem{WeidnerICRA2017}
N.~Weidner, S.~Rahman, A.~{Quattrini Li}, and I.~Rekleitis.
\newblock Underwater cave mapping using stereo vision.
\newblock In {\em IEEE International Conference on Robotics and Automation
  (ICRA)}, pages 5709 -- 5715, 2017.

\bibitem{zhang2018road}
Z.~Zhang, Q.~Liu, and Y.~Wang.
\newblock Road extraction by deep residual u-net.
\newblock {\em IEEE Geoscience and Remote Sensing Letters}, 15(5):749--753,
  2018.

\bibitem{zhang2018generalized}
Z.~Zhang and M.~Sabuncu.
\newblock Generalized cross entropy loss for training deep neural networks with
  noisy labels.
\newblock {\em Advances in Neural Information Processing Systems}, 31, 2018.

\bibitem{zhou2019unet++}
Z.~Zhou, M.~M.~R. Siddiquee, N.~Tajbakhsh, and J.~Liang.
\newblock Unet++: Redesigning skip connections to exploit multiscale features
  in image segmentation.
\newblock {\em IEEE transactions on medical imaging}, 39(6):1856--1867, 2019.

\bibitem{zhu2020saliency}
J.~Zhu, S.~Yu, L.~Gao, Z.~Han, and Y.~Tang.
\newblock {Saliency-Based Diver Target Detection and Localization Method}.
\newblock {\em {Mathematical Problems in Engineering}}, 2020, 2020.

\end{thebibliography}
